\documentclass[
 aps,
 prl,
 twocolumn,
 superscriptaddress,
 nofootinbib,
 showpacs,preprintnumbers,amsmath,amssymb,
 nofootinbib,
 longbibliography,
 floatfix,
 10pt
]{revtex4-2}

\usepackage{graphicx}
\usepackage{bm,latexsym,amsmath,amssymb,amsfonts,mathrsfs,textcomp}
\usepackage{color}
\input{colordvi.tex}
\usepackage{comment}
\usepackage{simpler-wick}
\usepackage{tikz}
\usepackage[compat=1.1.0]{tikz-feynhand}
\usepackage[normalem]{ulem}

\usepackage[
colorlinks=true, 
pdfstartview=FitV, 
linkcolor=blue, 
citecolor=magenta, 
urlcolor=blue, 
bookmarks=true,
bookmarksnumbered=true,
pdftitle={},
pdfauthor={}
]{hyperref}

\newcommand\sect[1]{{\it #1.}---}%Define section

\renewcommand\Im{\mathop{\mathrm{Im}}}

\newcommand{\rmd}{\mathrm d}
\newcommand{\rme}{\mathrm e}
\newcommand{\rmi}{\mathrm i}
\newcommand{\average}[1]{\langle #1\rangle}

\newcommand{\for}{\textrm{for}}

\newcommand{\eff}{\textrm{eff}}
\newcommand{\micro}{\mathrm{micro}}
\newcommand{\short}{\mathrm{short}}
\newcommand{\boundary}{\mathrm{bdy}}
\newcommand{\tail}{\mathrm{tail}}

\newcommand{\bx}{\bm{x}}

\newcommand{\bzero}{\bm{0}}

\newcommand{\bk}{\bm{k}}

\newcommand{\calD}{\mathcal{D}}

\newcommand{\calL}{\mathcal{L}}

\newcommand{\calT}{\mathcal{T}}

\begin{document}

\title{Universal Long-distance Tail of Particle-number Density near a Wall in Superfluids}

\author{Keisuke Fujii}
\email{fujii@phys.sci.isct.ac.jp}
\affiliation{Department of Physics, Institute of Science Tokyo,
Ookayama, Meguro-ku, Tokyo 152-8551, Japan}

\begin{abstract}
We investigate the long-distance recovery of the particle-number density near a static, impermeable wall in a $d$-dimensional Galilean-invariant superfluid.
Using superfluid effective field theory (EFT) on a half-space, we show that quantum phonon fluctuations produce a $z^{-(d+1)}$ tail in the smooth density profile at zero temperature, where $z$ is the distance from the wall.
The tail coefficient is fixed by the bulk equation of state and is independent of microscopic wall details.
At low temperature $T$, the thermal length $c_s/T$, with $c_s$ the bulk sound speed, separates the zero-temperature behavior at intermediate distances from a thermal $Tz^{-d}$ tail at larger distances for $d>1$.
For $d=1$, the long-distance phonon contribution instead decays as $\exp(-4\pi Tz/c_s)$.
The EFT applies to both Bose and Fermi superfluids, including strongly interacting Fermi gases at unitarity.
We estimate the normalized tail amplitude in three dimensions across the BCS--BEC crossover using experimental sound-speed data.
\end{abstract}

\maketitle

Boundaries are unavoidable when considering matter of finite extent.
They impose local constraints that can modify the spatial profiles of observables nonperturbatively, giving rise to phenomena absent in homogeneous systems~\cite{Diel:1997}.
The range of these changes is governed by low-energy excitations, with gapless modes allowing boundary effects to extend far into the interior.
Superfluids are paradigmatic systems with a gapless phonon associated with spontaneous breaking of the particle-number $U(1)$ symmetry~\cite{Goldstone:1962es,Hugenholtz:1959}.
Its presence raises the question of how phonon fluctuations affect the recovery of the density away from a boundary.

Cold-atom Bose and Fermi gases in optical box potentials provide an experimental setting for this question, combining a nearly uniform interior with steep confining walls~\cite{Meyrath:2005,Mukherjee:2017,Navon:2021}.
These traps have enabled studies of condensation and coherence in nearly uniform Bose gases in different dimensions~\cite{Gaunt:2013,Chomaz:2015,Navon:2015}.
Quantum fluctuations and collective dynamics have also been explored in this setting~\cite{Lopes:2017,Navon:2016,Ville:2018,Gauthier:2019,Christodoulou:2021,Hilker:2022}.

Theoretically, a superfluid near a sufficiently steep planar wall can be modeled locally on a half-space.
For a weakly interacting Bose gas with an impenetrable wall, the Gross--Pitaevskii mean-field solution is $n_{\mathrm{GP}}(z)=n_0\tanh^2[z/(\sqrt{2}\xi)]$, where $n_0$ is the bulk density, $\xi\equiv1/\sqrt{2m\mu}$ is the healing length, and $z$ is the distance from the wall~\cite{Pethick:2008}.
Far from the wall, the density recovers exponentially to its bulk value, $n_{\mathrm{GP}}(z)\simeq n_0[1-4\exp(-\sqrt{2}z/\xi)]$.

The absence of a long-range phonon contribution at mean-field level can be understood from the coupling to an external potential.
For a weakly interacting Bose gas, we write the microscopic field as $\psi=\sqrt{n_{\mathrm{GP}}(z)+\eta}\,\exp(i\varphi)$, where $\eta$ and $\varphi$ denote density and phase fluctuations, respectively.
At mean-field level, the static density response is screened on the healing-length scale, whereas the phase field carries the long-range response associated with the gapless phonon.
The leading long-wavelength effective coupling of an external potential that is linear in $\varphi$ involves $\partial_t\varphi$ and therefore generates no static long-range response.
The leading static phonon contribution instead arises from couplings quadratic in $\varphi$, through fluctuation averages beyond mean field~\cite{Fujii:2022,Akamatsu:2024}.
Reflection at the wall modifies these fluctuations far beyond the microscopic boundary layer, producing a long-distance correction to the mean density.

A similar distinction arises for heavy impurities weakly coupled to a weakly interacting Bose gas, where mean-field theory predicts an exponentially decaying Yukawa interaction~\cite{Nakano:2016,Naidon:2018,Camacho:2018PRL,Camacho:2018PRX}.
Quantum fluctuations of superfluid phonons generate a van der Waals interaction that dominates over the mean-field Yukawa potential at large separations~\cite{Fujii:2022}.
The impurity-medium coupling is treated perturbatively in this setting, whereas an impermeable wall requires an inherently nonperturbative treatment through the phonon boundary condition.

\begin{figure}[t]
 \centering
\includegraphics[width=0.7\linewidth]{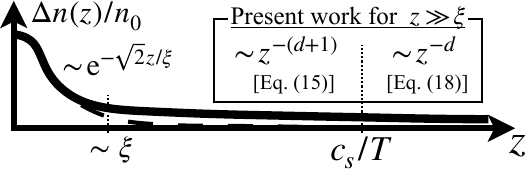}
\caption{Schematic density recovery for $d>1$, shown as $\Delta n(z)/n_0$, where $\Delta n(z)=n(\infty)-n(z)$.
 The exponential part (dashed line) illustrates the mean-field profile of a weakly interacting Bose gas with healing length $\xi$.
 Beyond the microscopic boundary layer, phonon fluctuations give $z^{-(d+1)}$ recovery for $z\ll c_s/T$ and $Tz^{-d}$ for $z\gg c_s/T$.
The curves illustrate the successive recovery regimes; their shapes, relative amplitudes, and crossover positions are schematic.
 }
 \label{fig:regimes}
\end{figure}

In this Letter, we derive a universal phonon contribution to density recovery near a wall from quantum fluctuations, which eventually dominates the exponential recovery at sufficiently large distances.
At zero temperature, the phonon contribution to the density difference $n(\infty)-n(z)$ generically scales as $z^{-(d+1)}$ in $d$ spatial dimensions.
At low nonzero temperatures, this contribution crosses over for $z\gg c_s/T$ to a thermal $Tz^{-d}$ tail for $d>1$ (see Fig.~\ref{fig:regimes}) and an exponential $\exp(-4\pi Tz/c_s)$ decay for $d=1$.
Here, $c_s$ is the bulk sound speed and $T$ is the temperature.
The leading tail is determined by the bulk equation of state, allowing the results to apply to both Bose and Fermi superfluids, including Fermi gases at unitarity.
We estimate its normalized zero-temperature amplitude in three dimensions across the BCS--BEC crossover using experimental sound-speed data~\cite{Hoinka:2017}.

We formulate the half-space theory by supplementing the bulk superfluid effective field theory (EFT)~\cite{Greiter:1989,SonWingate:2006} with a local boundary action that encodes unresolved short-distance physics near the wall~\cite{Burgess:2020,Cuomo:2021}.
Related treatments of short-distance boundary effects have been developed in EFT calculations of Casimir energies~\cite{Ravndal:2001,Aghababaie:2004}.
Within this formulation, we derive the leading Neumann condition for the continued bulk phase field.
Our result reproduces the zero-temperature tail in weakly interacting one-dimensional Bose gases~\cite{Reichert:2019wall,Petkovic:2020} and identifies its universal origin in phonon propagation and reflection.

\sect{Effective theory for superfluids on a half-space}
We consider a $d$-dimensional Bose gas or attractive Fermi gas in a box-shaped or sufficiently steep potential $V_{\mathrm{trap}}(\bx)$ that vanishes in the bulk.
The wall is assumed to be static, impermeable, and to preserve the particle-number $U(1)$ symmetry.
We model the region near a planar wall by a half-space, with the physical wall at $z=0$.
The gas occupies $z>0$, with vacuum on the other side of the wall; we write the spatial coordinates as $\bx=(\bx_{\parallel},z)$, where $\bx_{\parallel}\in\mathbb{R}^{d-1}$ denotes the coordinates parallel to the wall and $z>0$ is the distance from it.
For $d=1$, parallel derivatives vanish, and integrals over parallel coordinates or momenta are omitted.
We assume a superfluid ground state whose only independent gapless degree of freedom is the superfluid phonon.
In low dimensions, we apply the same EFT when coarse-grained density fluctuations and local phase gradients remain small, even if long-wavelength phase fluctuations destroy true long-range order~\cite{Mora:2003}.
For $d=1$, the density calculated below is its smooth phonon contribution; oscillatory density harmonics~\cite{Cazalilla:2004} are outside this local phase description.
We use natural units $\hbar=k_B=1$.

We first recall the bulk EFT for a homogeneous superfluid without a trapping potential.
At leading order in the gradient expansion, the EFT for the phase field $\varphi(t,\bx)$ has the Lagrangian density
\begin{align}
\calL_{\eff}(t,\bx)=p(\theta(t,\bx))\label{eq:EFT}
\end{align}
with $\theta(t,\bx)=\mu-\partial_{t}\varphi(t,\bx)-[\nabla\varphi(t,\bx)]^{2}/(2m)$.
Here, $p(\mu)$ is the pressure of the homogeneous system as a function of the chemical potential $\mu$, and $m$ is the particle mass.
At this order, particle-number symmetry and Galilean invariance constrain the phase dependence to the combination $\theta(t,\bx)$~\cite{Greiter:1989,SonWingate:2006}.
The bulk reference density and susceptibility are $n_0=p'(\mu)>0$ and $\chi=p''(\mu)>0$, respectively.

We then consider the EFT for the superfluid on a half-space.
Near the physical wall at $z=0$, the particle-number density microscopically varies on a short length scale $\ell_{\short}$, such as the healing length $\xi$, and vanishes at the wall.
These variations are not resolved by the bulk derivative expansion underlying Eq.~\eqref{eq:EFT}.
For $z\gg\ell_{\short}$, the same bulk EFT describes the long-wavelength phase field, which we call the outer field.

For $k\ell_{\short}\ll1$, we extrapolate the outer field to $z=0$ and formulate the half-space EFT in terms of this continued field $\varphi(t,\bx)$.
The boundary value $\varphi(t,\bx_\parallel,0)$ denotes the extrapolated outer field, rather than a microscopic phase at the wall, where the density vanishes.
The effects of the unresolved surface layer can then be represented by local boundary contributions, defined relative to this continued bulk description~\cite{Burgess:2020,Cuomo:2021}.
Including the boundary contribution, the half-space action can be written as $S_{\mathrm{tot}}[\varphi]=S_{\eff}[\varphi]+S_{\boundary}[\varphi]$.
Here $S_{\eff}$ denotes the leading bulk action given by
\begin{align}
S_{\eff}[\varphi]
=\int_{z>0}\rmd t\,\rmd^d\bx\,p(\theta(t,\bx)).
\label{eq:LO-effective-action}
\end{align}
The boundary action $S_{\boundary}$ is a local functional at $z=0$, expanded in time derivatives and derivatives parallel to the wall.

We show that $S_{\boundary}$ is subleading in the gradient expansion for fixed, finite boundary coefficients, so that the leading order effective action is given by $S_{\eff}[\varphi]$ in Eq.~\eqref{eq:LO-effective-action}.
We use the counting $\partial\varphi=O(1)$, in which each additional derivative raises the order by one~\cite{SonWingate:2006}.
To compare the boundary contribution with the bulk Lagrangian density, we express
$S_{\boundary}[\varphi]=\int_{z\geq0}\rmd t\,\rmd^d\bx\,\delta_+(z)\calL_{\short}$,
where $\delta_+(z)$ is normalized by $\int_0^\infty\rmd z\,\delta_+(z)f(z)=f(0)$.
Although $p(\theta)$ and $\calL_{\short}$ are both $O(1)$, the boundary contribution contains the additional factor $\delta_+(z)$.
To see this, we introduce the dimensionless coordinate $Z=kz$, which measures the distance from the wall in units of $k^{-1}$.
In this coordinate, the boundary delta function becomes $\delta_+(z)=k\,\delta_+(Z)$.
For fields varying over distances of order $k^{-1}$, this factor makes the boundary action one order higher than the bulk action.

\sect{Effective boundary condition}
We use $S_{\eff}[\varphi]$ in Eq.~\eqref{eq:LO-effective-action} alone for the leading-order analysis below.
Since the wall does not fix the phase, we vary the continued field $\varphi$, including its boundary value:
\begin{align}
\delta S_{\eff}
=\int_{z>0}\!\!\!\rmd t\,\rmd^d\bx\,
(\partial_t n+\bm\nabla\cdot\bm j)\delta\varphi
+\int_{z=0}\!\!\!\rmd t\,\rmd^{d-1}\bx_\parallel\,j_z\delta\varphi.
\label{eq:bulk-action-variation}
\end{align}
Here, $n(t,\bx)=p^{\prime}(\theta(t,\bx))$ and $\bm{j}(t,\bx)=n(t,\bx)\nabla\varphi(t,\bx)/m$ are the particle-number density and current in the bulk EFT, respectively.
Bulk variations give the continuity equation $\partial_t n+\bm\nabla\cdot\bm j=0$.

The boundary term in Eq.~\eqref{eq:bulk-action-variation} must then vanish for arbitrary $\delta\varphi|_{z=0}$, giving the no-flux condition $j_z|_{z=0}=0$ at this order.
At linear order in the fluctuations, the current in the $z$-direction is $j_{z}(t,\bx)=n_{0}\partial_{z}\varphi(t,\bx)/m+O((\partial\varphi)^{2})$.
The bulk density relation $n=p'(\theta)$ is evaluated on the continued outer field, so that in the homogeneous reference state with $\theta=\mu$, its value is $n_0$ even at $z=0$.
Dividing this leading EFT no-flux condition by the nonzero coefficient $n_0/m$ of its linearized current gives the Neumann boundary condition,
\begin{align}
\partial_{z}\varphi(t,\bx)|_{z=0}=0.
\label{eq:Neumann}
\end{align}
It should be emphasized that this division acts on the continued EFT current, not on the exact microscopic no-flux condition at the wall, where the density vanishes.

We examine boundary corrections to illustrate explicitly their subleading order in the gradient expansion.
At finite frequency or wave number, the surface layer can exchange particles with the bulk and carry tangential current, producing corrections through the boundary action.
To obtain the correction to the linear phonon boundary condition, we retain the leading local terms in $S_{\boundary}$ quadratic in fluctuations:
\begin{align}
S_{\boundary}^{(2)}[\varphi]
=\int_{z=0}\rmd t\rmd^{d-1}\bx_{\parallel}\,
\biggl[
	\frac{g^{t}}{2}(\partial_{t}\varphi)^{2}+\frac{g^{s}}{2}(\nabla_{\parallel}\varphi)^{2}
\biggr].
\end{align}
Here $g^t$ and $g^s$ are real coefficients that remain finite in the low-frequency, long-wavelength limit.
Adding its variation to Eq.~\eqref{eq:bulk-action-variation} gives the boundary condition from $S_{\mathrm{tot}}$ at linear order in $\varphi$:
\begin{align}
\biggl[\frac{n_{0}}{m}\partial_{z}-g^{t}\partial^{2}_{t}-g^{s}\nabla_{\parallel}^{2}\biggr]\varphi|_{z=0}=0.
\label{eq:corrected-boundary-condition}
\end{align}
In the counting $\partial\varphi=O(1)$, the last two terms contain one additional derivative relative to the bulk normal current and therefore give next-to-leading corrections for finite $g^t$ and $g^s$.
In the Supplemental Material~\cite{SM}, we determine these corrections for a three-dimensional weakly interacting Bose gas by matching the microscopic solution to the outer phonon solution in the overlap region $\xi\ll z\ll k^{-1}$.
This microscopic matching yields the effective boundary condition~\eqref{eq:corrected-boundary-condition} with $g^{t}=-\chi\xi/\sqrt{2}$ and $g^{s}=\sqrt{2}n_{0}\xi/m$.

\sect{Long-distance tail at zero temperature}
We calculate the leading quantum correction to the ground-state particle-number density at long distances $z\gg\ell_{\short}$, using the leading Neumann boundary condition~\eqref{eq:Neumann}.
Expanding the bulk density to quadratic order in the phonon fluctuations gives
\begin{align}
n(t,\bx)&=n_{0}-\chi\partial_{t}\varphi(t,\bx) \nonumber \\
&\quad
-\frac{\chi}{2m}(\nabla\varphi)^{2}+\frac{\chi^{\prime}}{2}(\partial_{t}\varphi)^{2}+O((\partial\varphi)^{3}),\label{eq:expanded-density}
\end{align}
with $\chi'=\partial_\mu\chi$.
Taking the ground-state average, we define $\delta n(z)$ relative to the homogeneous reference density $n_0$ by $\average{n(z)}=n_0+\delta n(z)$.
Stationarity gives $\average{\partial_t\varphi}=0$, so that Eq.~\eqref{eq:expanded-density} yields
\begin{align}
\delta n(z)
=-\frac{\chi}{2m}\average{(\bm\nabla\varphi)^2}
+\frac{\chi'}{2}\average{(\partial_t\varphi)^2}+O(\varphi^{3}).
 \label{eq:density-contractions}
\end{align}
At leading one-loop order, these averages are evaluated in the Gaussian phonon theory, in which odd moments, including the cubic density terms, vanish.

Expanding Eq.~\eqref{eq:LO-effective-action} to quadratic order in the phonon fluctuations gives
\begin{align}
S_{\eff}
&=\int_{z>0} \rmd t\rmd^{d}\bx\,
\biggl[
	\frac{1}{2}(\partial_{t}\pi)^{2}-\frac{c_{s}^{2}}{2}(\nabla\pi)^{2}
\biggr]+O(\pi^{3}),
\label{eq:expanded-action}
\end{align}
where $\pi=\sqrt{\chi}\varphi$ is the canonically normalized field and $c_{s}\equiv \sqrt{n_{0}/(m\chi)}$ is the bulk sound speed.
We introduce the phonon propagator, $\rmi G(t,\bx;t^{\prime},\bx^{\prime})=\average{\calT\pi(t,\bx)\pi(t^{\prime},\bx^{\prime})}$, where $\calT$ denotes time ordering.
The propagator satisfies the Neumann condition~\eqref{eq:Neumann} in either normal coordinate, with the other held fixed in $z>0$.

The modes $\sqrt{2/\pi}\cos(k_z z)$ are normalized on the half-line $z>0$ with the Neumann boundary condition, giving the propagator
\begin{align}
G(\omega,\bk_{\parallel};z,z^{\prime})
=\frac{2}{\pi}\int^{\infty}_{0}\rmd k_{z}\, G_{\mathrm{b}}(\omega,\bk)\cos(k_{z}z)\cos(k_{z}z^{\prime}),
\end{align}
where $G(\omega,\bk_\parallel;z,z^{\prime})$ is the Fourier transform in time and the coordinates parallel to the wall.
Here, $G_{\mathrm{b}}(\omega,\bk)$ is the propagator in the infinite homogeneous bulk,
\begin{align}
G_{\mathrm{b}}(\omega,\bk)=\frac{1}{\omega^{2}-c_{s}^{2}k^{2}+\rmi 0^{+}}
\end{align}
with $\bk=(\bk_\parallel,k_z)$ and $k^2=|\bk_\parallel|^2+k_z^2$.
Using $2\cos(k_{z}z)\cos(k_{z}z^{\prime})=\cos[k_{z}(z-z^{\prime})]+\cos[k_{z}(z+z^{\prime})]$, we can decompose the propagator as
\begin{align}
G(\omega,\bk_{\parallel};z,z^{\prime})=
G_{\mathrm{b}}(\omega,\bk_{\parallel};z-z^{\prime})
+G_{\mathrm{b}}(\omega,\bk_{\parallel};z+z^{\prime})\label{eq:image-decomposition}
\end{align}
with $G_{\mathrm{b}}(\omega,\bk_{\parallel};z)
=\int^{\infty}_{-\infty}\rmd k_{z}/(2\pi)\,G_{\mathrm{b}}(\omega,\bk)\rme^{\rmi k_{z}z}$.
The second term is the image of the bulk propagator under $z'\to-z'$, implementing reflection at the wall.

The coincident-point contractions give
\begin{align}
\delta n(z)=\lim_{t^{\prime}\to t^+}\lim_{\bx^{\prime}\to \bx}
\biggl[
	-\frac{\nabla_{\bx}\!\cdot\!\nabla_{\bx^{\prime}}}{2m}
	+\frac{\chi^{\prime}}{2\chi}\partial_{t}\partial_{t^{\prime}}
\biggr]\rmi G(t,\bx;t^{\prime},\bx^{\prime}).\label{eq:limit-dn}
\end{align}
The direct and image terms in Eq.~\eqref{eq:image-decomposition} give the decomposition
\begin{align}
\delta n(z)=\delta n_{\mathrm{b}}+\delta n_{\mathrm{tail}}(z).
\end{align}
The direct term gives the $z$-independent homogeneous contribution $\delta n_{\mathrm b}$, whose renormalization is included in matching the bulk equation of state.
It cancels in the density difference $\average{n(z)}-\average{n(\infty)}=\delta n_{\tail}(z)$.

The image term, evaluated at $z+z'=2z$ in the coincident-point limit, gives
\begin{align}
\delta n_{\tail}(z)
&=-\frac{\Gamma((d+1)/2)}
{2^{d+3}\pi^{(d+1)/2}z^{d+1}}
\left(\frac{2d-1}{m c_s}+c_s\frac{\chi'}{\chi}\right).
\label{eq:general_zeroT}
\end{align}
The regularization and evaluation of the coincident-point contractions are described in the Supplemental Material~\cite{SM}.
The two derivatives in the density operator turn the massless propagator's $z^{-(d-1)}$ behavior for $d>1$, or logarithmic behavior for $d=1$, into the $z^{-(d+1)}$ tail.
Additional phonon loops are suppressed at long distances, so that Eq.~\eqref{eq:general_zeroT} gives the leading nonzero tail when its coefficient does not vanish.

For a weakly interacting one-dimensional Bose gas with contact coupling $g_{\mathrm{1D}}$, the leading equation of state $p(\mu)=\mu^2/(2g_{\mathrm{1D}})$ gives $\chi'=0$.
Equation~\eqref{eq:general_zeroT} then gives $\delta n_{\tail}(z)=-1/(16\pi m c_s z^2)$, reproducing the microscopic zero-temperature result, including its coefficient~\cite{Reichert:2019wall,Petkovic:2020}.

\sect{Finite-temperature effect}
We calculate the finite-temperature phonon contribution using the Matsubara formalism~\cite{Matsubara:1955,AGD:1959}.
We work at $T\ell_{\short}/c_s\ll1$, with thermal excitations dominated by phonons, and retain the zero-temperature bulk coefficients and the leading Neumann boundary condition.
The temperature $T$ introduces a thermal length scale $c_s/T$, and the behavior of the tail changes depending on the relative scales of this length and the distance $z$.
Replacing the ground-state average by the equilibrium average, we write $\average{n(z)}_T=n_0+\delta_T n(z)$ with
\begin{align}
\delta_{T}n(z)=-\frac{\chi}{2m}\average{(\bm\nabla\varphi)^2}_{T}
+\frac{\chi'}{2}\average{(\partial_t\varphi)^2}_{T},
\end{align}
where $\average{\cdots}_T$ denotes the equilibrium average at temperature $T$.
As at zero temperature, we separate the direct and image contributions, $\delta_T n(z)=\delta_T n_{\mathrm b}+\delta_T n_{\tail}(z)$.
The homogeneous term $\delta_T n_{\mathrm b}$ contains the bulk thermal shift, while $\delta_T n_{\tail}(z)$ is the wall-induced density difference from the bulk at the same temperature and chemical potential.

With $t=-\rmi\tau$, we introduce the Matsubara propagator $\Delta=\average{\calT_\tau\pi(\tau,\bx)\pi(\tau^{\prime},\bx^{\prime})}_T$ with imaginary-time ordering $\calT_\tau$.
For Matsubara frequencies $\omega_j=2\pi jT$ with $j\in\mathbb Z$, its image part is
$\Delta(\omega_j,\bk_\parallel;z+z^{\prime})=\rme^{-Q_j(z+z^{\prime})}/(2c_s^2Q_j)$
with $Q_j=\sqrt{k_\parallel^2+\omega_j^2/c_s^2}$.
Applying the density operator in Eq.~\eqref{eq:limit-dn} with $\partial_t=\rmi\partial_\tau$ and taking derivatives before the limit gives
\begin{align}
\delta_T n_{\tail}(z)
&=-T\sum_{j}
\int\frac{\rmd^{d-1}\bk_\parallel}{(2\pi)^{d-1}}
\frac{\rme^{-2Q_jz}}{2c_{s}^{2}Q_j} \nonumber \\
&\quad\times
\left[
	\frac{k_\parallel^2+Q_j^2}{2m}
	+\frac{\chi'}{2\chi}\omega_j^2
\right].
\label{eq:thermal-tail-sum}
\end{align}

For $\ell_{\short}\ll z\ll c_s/T$, the leading term reduces to the zero-temperature tail~\eqref{eq:general_zeroT}.
For $d>1$ and $z\gg c_s/T$, the Matsubara zero mode $\omega_j=0$ gives the dominant contribution and the density deviation is obtained as
\begin{align}
\delta_T n_{\tail}(z)
\simeq-\frac{(d-1)\Gamma(d/2)}
{2^{d+1}\pi^{d/2}}\frac{T}{m c_s^2z^d}.
\label{eq:thermal-tail-asymptote}
\end{align}
For $d=1$, the zero-mode image Green function is linear in $z+z'$ up to an infrared-divergent constant, so that its density contribution vanishes for $z>0$.
The nonzero Matsubara modes instead give, at the leading phonon order considered here,
\begin{align}
\left.\delta_T n_{\tail}(z)\right|_{d=1}
=-\frac{\pi T^2}{4c_s^2\sinh^2(2\pi Tz/c_s)}
\left(\frac{1}{mc_s}+c_s\frac{\chi'}{\chi}\right).
\label{eq:thermal-tail-1D}
\end{align}

\sect{Amplitude of the density tail}
We now set $d=3$ and estimate the magnitude of the zero-temperature tail in Eq.~\eqref{eq:general_zeroT}.
It is worth emphasizing that the tail is universal, independent of microscopic details, and the bulk input to its coefficient consists only of the sound speed and its derivative.
The asymptotic result applies to Bose superfluids and to Fermi superfluids throughout the BCS--BEC crossover, including the unitary Fermi gas.
Practically, the $\chi^{\prime}$ term in Eq.~\eqref{eq:general_zeroT} scales as $mc_{s}^{2}\chi^{\prime}/\chi\sim (c_{s}/v_{F})^{2}$, so that we ignore it in the estimation below.
With $n_0=k_F^3/(3\pi^2)$ and $\xi=1/(\sqrt{2}mc_s)$, this estimate gives $|\delta n_{\tail}(z)|/n_0\simeq(15/16)(c_s/v_F)^3(\xi/z)^4$.

\begin{figure}[t]
 \centering
 \includegraphics[width=0.88\linewidth]{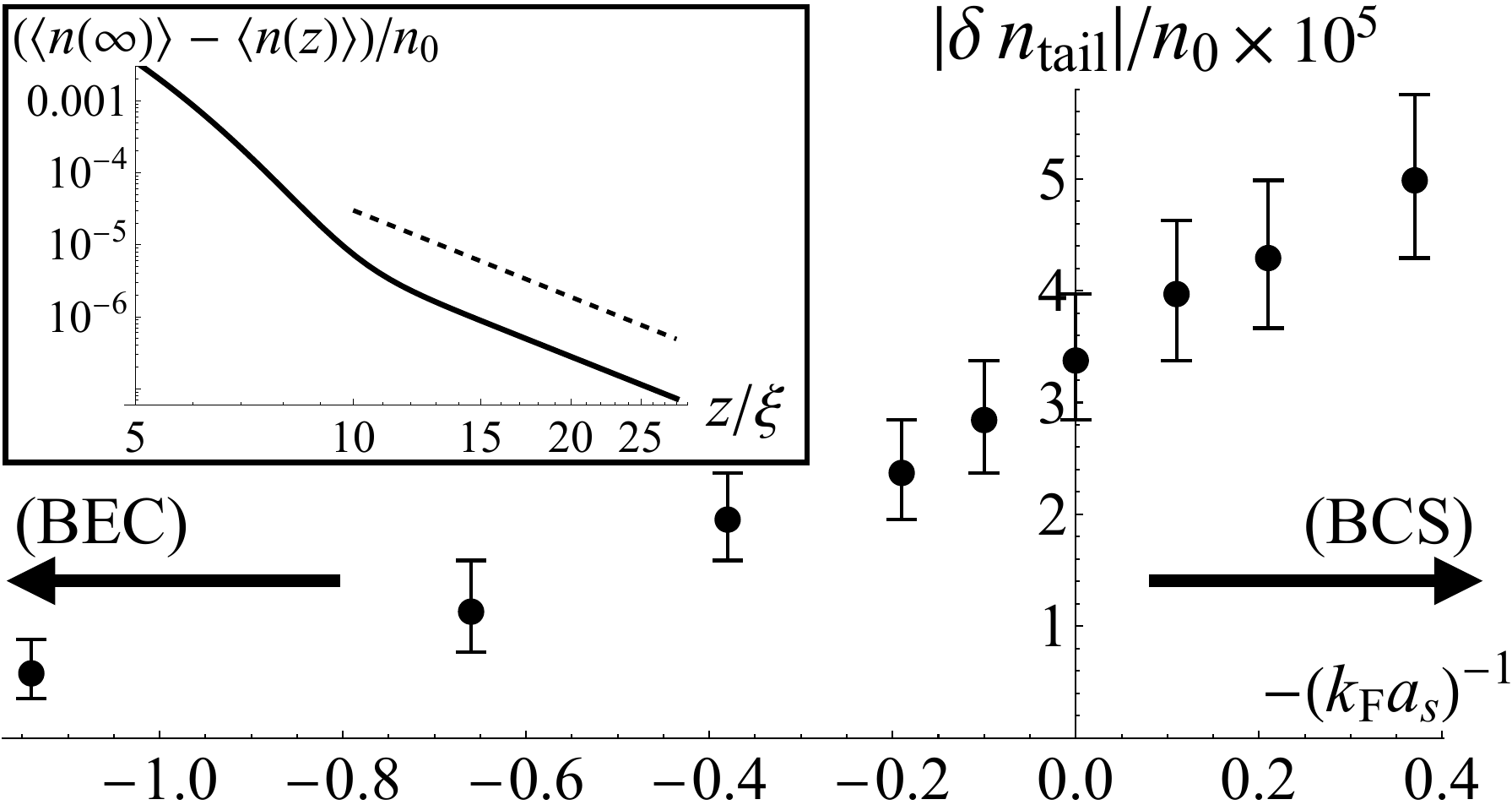}
\caption{
Normalized asymptotic tail, $|\delta n_{\tail}(z)|/n_0$, evaluated at the reference distance $z=6\xi$ across the BCS--BEC crossover.
The points use the measured $c_s/v_F$ of Ref.~\cite{Hoinka:2017} and neglect the $\chi'$ contribution in Eq.~\eqref{eq:general_zeroT}.
Inset: double-logarithmic plot of the normalized density difference in Eq.~\eqref{eq:Bose-exponential-and-tail} for a dilute Bose gas with $n_0a_s^3=10^{-3}$ (solid line).
The dashed line indicates the $(z/\xi)^{-4}$ dependence, shifted vertically for clarity.
The two terms in Eq.~\eqref{eq:Bose-exponential-and-tail} become equal at $z\simeq9.6\xi$, beyond which the recovery approaches the $z^{-4}$ behavior.
}
\label{fig:plot}
\end{figure}

We can quantify this ratio by providing input parameters from, for example, experimental data of the sound speed~\cite{Hoinka:2017} or microscopic theoretical calculations~\cite{Marini:1998,Roberto:2008,Schakel:2011,Klimin:2014}.
Figure~\ref{fig:plot} evaluates this asymptotic estimate at the reference distance $z=6\xi$ as a function of $-(k_Fa_s)^{-1}$, where $a_s$ is the fermionic $s$-wave scattering length.
At fixed $z/\xi$, the estimated normalized tail increases toward the BCS side over the range shown.

For the dilute three-dimensional Bose gas shown in the inset, we use $p(\mu)=\mu^2/(2g)$ with $g=4\pi a_s/m$ and $n_0a_s^3\ll1$, where $a_s$ is the bosonic $s$-wave scattering length.
For $z\gg\xi$, combining the tree-level exponential recovery with the leading power-law tail gives
\begin{equation}
\frac{\average{n(\infty)}-\average{n(z)}}{n_0}
\simeq4\rme^{-\sqrt{2}z/\xi}+\frac{5}{2\sqrt{\pi}}\sqrt{n_0a_s^3}
\left(\frac{\xi}{z}\right)^4.
\label{eq:Bose-exponential-and-tail}
\end{equation}

\sect{Discussion and outlook}
In this Letter, we have derived the long-distance power-law tail of the particle-number density near a planar wall in a Galilean-invariant superfluid in $d$ spatial dimensions.
The wall-induced phonon contribution decays as $z^{-(d+1)}$ at zero temperature [Eq.~\eqref{eq:general_zeroT}] and as $Tz^{-d}$ for $d>1$ at $z\gg c_s/T$ [Eq.~\eqref{eq:thermal-tail-asymptote}].
For $d=1$, a nonzero smooth phonon contribution instead decays as $\exp(-4\pi Tz/c_s)$ at the order considered here [Eq.~\eqref{eq:thermal-tail-1D}].
Although its small estimated amplitude may make experimental detection in box traps challenging (Fig.~\ref{fig:plot}), the zero-temperature power-law tail is a universal consequence of gapless phonon fluctuations and yields asymptotic density recovery qualitatively distinct from the mean-field exponential profile.

Universality follows from long-wavelength phonon propagation and reflection with the leading Neumann condition imposed on the continued bulk field.
It should be emphasized that this effective no-flux condition is a leading-order statement in the gradient expansion, distinct from the exact microscopic no-flux condition at the physical wall.
The division used to obtain the Neumann condition is by the nonzero bulk density $n_0$ in the linearized EFT current, not by the microscopic density that vanishes at the wall.
Higher-order boundary responses modify this effective condition, while the exact microscopic no-flux condition at the physical wall remains unchanged.
The Supplemental Material~\cite{SM} determines these corrections~\eqref{eq:corrected-boundary-condition} by microscopic matching for a three-dimensional weakly interacting Bose gas.

Our treatment of a rigid wall provides a starting point for studying other local observables, including the energy density~\cite{Henkel:2026border}, and for treating more general boundaries and localized defects.
Effective theories for superfluid interfaces and self-bound droplets already retain the interface displacement as a dynamical field coupled to the bulk phonon~\cite{Watanabe:2014,Armas:2017,Mitsuhashi:2026}.
These settings raise the question of which features of the low-energy response are fixed by symmetry and bulk properties, and which depend on interface or defect parameters.
Conformal symmetry provides additional constraints on effective descriptions of boundaries and defects in relativistic field theories~\cite{Cuomo:2021,Diatlyk:2024,Kravchuk:2024}.
For the unitary Fermi gas, nonrelativistic conformal symmetry constrains effective descriptions of trapped-cloud edges~\cite{Orland:2021,Hellerman:2021,Pellizzani:2022,Hellerman:2024,Pellizzani:2025PhDthesis}.
Adapting this approach to the response at a rigid wall is a direction for future work.

\begin{acknowledgments}
K.F. is supported by JSPS KAKENHI Grant Numbers JP24KJ0062 and JP26K17064.
The author thanks Kazuya Fujimoto and especially Yusuke Nishida for discussions.
\end{acknowledgments}

\bibliography{superfluid-in-box}

%---------   Supplemental Material   ---------%
%%%%%%%%%% Merge with supplemental materials %%%%%%%%%%
\appendix
\pagebreak
\widetext
\begin{center}
\textbf{\large Supplemental Material:\\
Universal Long-distance Tail of Particle-number Density near a Wall in Superfluids
}
\end{center}
%%%%%%%%%% Merge with supplemental materials %%%%%%%%%%
%%%%%%%%%% Prefix a "S" to all equations, figures, tables and reset the counter %%%%%%%%%%
\setcounter{equation}{0}
\setcounter{figure}{0}
\setcounter{table}{0}
\setcounter{page}{1}
\newcounter{supplementeqcountar}
%\makeatletter
\newcommand\suppsect[1]{{\it #1.}---}%Define section
\renewcommand{\theequation}{S\arabic{equation}}
\renewcommand{\thefigure}{S\arabic{figure}}
% \renewcommand{\bibnumfmt}[1]{[S#1]}
% \renewcommand{\citenumfont}[1]{S#1}
%%%%%%%%%% Prefix a "S" to all equations, figures, tables and reset the counter %%%%%%%%%%

%%%%%%%%%%%%%%%%%%%%%%%%%%%%%%%%%%%%%%%%%%%%%%%%%%%%%%%%%%%%%%%%%%%%%%%%%%
%%%%%%%%%%%%%%%%%%%%%%%%%%%%%%%%%%%%%%%%%%%%%%%%%%%%%%%%%%%%%%%%%%%%%%%%%%
%%%%%%%%%%%%%%%%%%%%%%%%%%%%%%%%%%%%%%%%%%%%%%%%%%%%%%%%%%%%%%%%%%%%%%%%%%
%%%%%%%%%%%%%%%%%%%%%%%%%%%%%%%%%%%%%%%%%%%%%%%%%%%%%%%%%%%%%%%%%%%%%%%%%%
\section{Microscopic calculation for a weakly interacting Bose gas}
\label{app-sect:weak-Bose-gas}

In this section, we retain density and phase fluctuations to derive the effective phonon boundary condition and check the long-distance density tail for a three-dimensional weakly interacting Bose gas.
At tree level in the homogeneous bulk, integrating out the density fluctuations at leading gradient order recovers the phase-only action used in the main text, with $p(\theta)=\theta^2/(2g)$.
The density-phase representation therefore provides a direct link between the microscopic density calculation and the phonon description of long-distance dynamics.

\subsubsection{Microscopic theory and tree-level density profile}
We start from the action
\begin{align}
S_{\micro}
=\int_{z>0} \rmd t\rmd^{3}\bx\,\biggl[
	\psi^{\dagger}\biggl(\rmi\partial_t+\frac{1}{2m}\nabla^2+\mu\biggr)\psi
	-\frac{g}{2}\bigl(\psi^{\dagger}\psi\bigr)^2
\biggr],
\end{align}
where $\psi$ is the bosonic field, $m$ is the particle mass, and $\mu$ is the chemical potential.
We impose $\psi|_{z=0}=0$ at the microscopic wall and restrict the spatial integration to $z>0$.
We assume a weak repulsive contact interaction with $g=4\pi a_s/m>0$, where $a_s$ is the $s$-wave scattering length.
Here $g$ is the renormalized contact coupling, with counterterms understood in loop calculations.
Keeping both density and phase degrees of freedom explicitly, we parameterize the field for $z>0$ as
\begin{align}
\psi(t,\bx)=\sqrt{n(t,\bx)}\, \rme^{\rmi \varphi(t,\bx)},\label{app-eq:embedding}
\end{align}
where $n(t,\bx)$ and $\varphi(t,\bx)$ are the density and phase fields, respectively.
Substituting this parameterization into the action gives
\begin{align}
S_{\micro}
&=
\int_{z>0} \rmd t\rmd^{3}\bx\,
\biggl[
	\mu n
	-\frac{g}{2}n^2
	-\frac{(\nabla \sqrt{n})^2}{2m}
	-n
	\biggl(
		\partial_t\varphi
		+\frac{\bigl(\nabla\varphi\bigr)^2}{2m}
	\biggr)
\biggr],
\label{app-eq:expanded-L}
\end{align}
where we dropped total derivative terms.

For a static density profile with constant phase, we take the variation of the action~\eqref{app-eq:expanded-L} as
\begin{align}
\left.\delta S_{\micro}\right|_{\varphi=\mathrm{const.}}
=\int_{z>0}\rmd t\,\rmd^3\bx\,
\biggl[\mu-gn+\frac{\partial_z^2\sqrt n}{2m\sqrt n}\biggr]\delta n
+\int\rmd t\,\rmd^2\bx_\parallel\,
\lim_{z\to0^+}\frac{\partial_z\sqrt n}{m}\,\delta\sqrt n.
\end{align}
The variation satisfies $\delta\sqrt n|_{z=0}=0$ because of the microscopic boundary condition $\delta\psi|_{z=0}=0$, so that the boundary term vanishes for finite $\partial_z\sqrt n$.
Stationarity in the bulk then gives
\begin{align}
0
=\left.\frac{\delta S_{\micro}}{\delta n}\right|_{\varphi=\mathrm{const.}}
=
\mu-g n(z)
 +\frac{n''(z)}{4m n(z)}
 -\frac{\bigl(n'(z)\bigr)^2}{8m\bigl(n(z)\bigr)^2},
 \qquad\for\qquad z>0,
 \label{app-eq:tree-density-equation}
\end{align}
with $n^{\prime}(z)\equiv\partial_{z}n(z)$ and $n^{\prime\prime}(z)\equiv\partial_{z}^{2}n(z)$.

In an infinite homogeneous system, the tree-level density is uniform, so that Eq.~\eqref{app-eq:tree-density-equation} gives $n_0=\mu/g$ for $\mu>0$.
We work in the dilute regime $n_0a_s^3\ll1$, which controls the weak-coupling loop expansion.
For the half-space geometry, we impose the boundary conditions
\begin{align}
n(0)=0,\qquad  n(z\to\infty)=n_{0}.
\end{align}
The positive ground-state solution $n_{\mathrm{GP}}(z)$ for $z>0$ is
\begin{align}
 n_{\mathrm{GP}}(z)=n_0\tanh^2\!\biggl(\frac{z}{\sqrt{2}\xi}\biggr),
 \qquad n_0=\frac{\mu}{g},
 \label{app-eq:tree-density-profile}
\end{align}
where $\xi\equiv1/\sqrt{2m\mu}$ is the healing length.
At tree level, the density recovers exponentially to its bulk value, with the asymptotic expansion for $z\gg\xi$
\begin{align}
n_{\mathrm{GP}}(z)= n_{0}\Bigl[1-4\rme^{-\sqrt{2}z/\xi}+O((\rme^{-\sqrt{2}z/\xi})^{2})\Bigr].
\end{align}

\subsubsection{Boundary conditions at the wall}
We first identify the microscopic boundary conditions on the fluctuations, which are used to construct the propagators and to perform the microscopic matching below.
We introduce the density fluctuation $\eta(t,\bx)$ on top of the tree-level profile through
\begin{align}
 n(t,\bx)=n_{\mathrm{GP}}(z)+\eta(t,\bx),
\end{align}
and set the constant background phase to zero, so that $\varphi(t,\bx)$ itself describes the phase fluctuation.

The microscopic boundary condition $\psi|_{z=0}=0$ requires $n|_{z=0}=0$, but does not fix the phase, which is undefined where the density vanishes.
To first order in the fluctuations, the deviation $\delta\psi(t,\bx)\equiv\psi(t,\bx)-\sqrt{n_{\mathrm{GP}}(z)}$ is
\begin{align}
\delta\psi(t,\bx)\simeq\frac{\eta(t,\bx)}{2\sqrt{n_{\mathrm{GP}}(z)}}
+\rmi\sqrt{n_{\mathrm{GP}}(z)}\,\varphi(t,\bx).
\label{app-eq:microscopic-fluctuation-boundary}
\end{align}
The real and imaginary parts of $\delta\psi|_{z=0}=0$ require
\begin{align}
\lim_{z\to0^+}\frac{\eta(t,\bx)}{2\sqrt{n_{\mathrm{GP}}(z)}}=0,\qquad
\lim_{z\to0^+}\sqrt{n_{\mathrm{GP}}(z)}\,\varphi(t,\bx)=0.\label{app-eq:boubdaey-cond}
\end{align}
Since $n_{\mathrm{GP}}(z)$ behaves near the wall as
\begin{align}
n_{\mathrm{GP}}(z)\sim\frac{n_0}{2\xi^2}z^{2},\qquad (z\to0^+)
\end{align}
smooth microscopic fluctuations satisfying $\delta\psi|_{z=0}=0$ obey
\begin{align}
\eta(t,\bx)=O(z^2),\qquad \varphi(t,\bx)=O(1)
\qquad (z\to0^+).
\label{app-eq:regular-wall-modes}
\end{align}
For these fluctuations, the microscopic current $j_z=\Im(\psi^*\partial_z\psi)/m$ vanishes at the wall.

\subsubsection{Fluctuations and their propagators}
After dropping the field-independent background terms and temporal total derivative terms, and using the background equation of motion to cancel linear terms, Eq.~\eqref{app-eq:expanded-L} becomes
\begin{align}
 S_{\micro}=
 \int_{z>0}\rmd t\,\rmd^3\bx\,
 \bigl[\calL_{\mathrm{kin}}+\calL_{\mathrm{int}}\bigr].
\end{align}
Here, $\calL_{\mathrm{kin}}$ contains all terms quadratic in $\eta$ and $\varphi$, and $\calL_{\mathrm{int}}$ starts at cubic order.
They are given by
\begin{align}
 \calL_{\mathrm{kin}}
 &=-\eta\partial_t\varphi
 -\frac{n_{\mathrm{GP}}(z)}{2m}(\nabla\varphi)^2
 -\frac{g}{2}\eta^2
 -\frac{n_{\mathrm{GP}}(z)}{8m}\Bigl(\nabla [\eta/n_{\mathrm{GP}}(z)]\Bigr)^2,
 \label{app-eq:inhomogeneous-quadratic-L}\\
 \calL_{\mathrm{int}}
 &=-\frac{\eta}{2m}(\nabla\varphi)^2
 +\frac{\eta}{8m\bigl(1+\eta/n_{\mathrm{GP}}(z)\bigr)}\Bigl(\nabla [\eta/n_{\mathrm{GP}}(z)]\Bigr)^2.
 \label{app-eq:inhomogeneous-interaction-L}
\end{align}
The interaction term retains all higher powers of $\eta/n_{\mathrm{GP}}(z)$ for $n_{\mathrm{GP}}+\eta>0$, while the loop expansion is organized in small fluctuation amplitudes about the GP background.
For the fluctuations in Eq.~\eqref{app-eq:regular-wall-modes}, $\eta/n_{\mathrm{GP}}$ and $\varphi$ remain bounded as $z\to0^+$, so that the vanishing background density does not by itself invalidate this amplitude expansion.

To find equations of motion, we take the variation of the quadratic action as
\begin{align}
\delta S_{\micro}^{(2)}
&=\int_{z>0}\rmd t\,\rmd^3\bx\,
\bigl[
(-\partial_t\varphi-\calD^{n}\eta)\delta\eta
+(\partial_t\eta-\calD^{\varphi}\varphi)\delta\varphi
\bigr] \nonumber \\
&\quad+\int\rmd t\,\rmd^2\bx_\parallel\,
\lim_{z\to0^+}\left[
\frac{n_{\mathrm{GP}}(z)}{m}\partial_z\varphi\,\delta\varphi
+\frac{1}{4m}\partial_z\!\left(\frac{\eta}{n_{\mathrm{GP}}(z)}\right)\delta\eta
\right].
\end{align}
Here, we define the spatial operators
\begin{align}
\calD^{n} f
&\equiv g f-\frac{1}{4mn_{\mathrm{GP}}(z)}\nabla\cdot\biggl[n_{\mathrm{GP}}(z)\nabla\biggl(\frac{f}{n_{\mathrm{GP}}(z)}\biggr)\biggr],
\label{app-eq:density-operator} \\
\calD^{\varphi}f
&\equiv -\frac{1}{m}\nabla\cdot \Bigl(n_{\mathrm{GP}}(z)\nabla f\Bigr),
\label{app-eq:phase-operator}
\end{align}
where $f$ is the function on which the operators act.
For variations preserving the boundary condition $\psi(0)=0$, $\delta\eta=O(z^2)$ and $\delta\varphi=O(1)$ hold for $z\to0^+$ as in Eq.~\eqref{app-eq:regular-wall-modes}.
The boundary term therefore vanishes, and setting the coefficients of the independent bulk variations to zero yields
\begin{align}
 -\partial_t\varphi-\calD^{n}\eta&=0,
 \label{app-eq:density-linear-eom}\\
 \partial_t\eta-\calD^{\varphi}\varphi&=0.
 \label{app-eq:phase-linear-eom}
\end{align}

We define the real-time propagators by
\begin{align}
 \rmi G_{ab}(t,\bx;t^{\prime},\bx^{\prime})\equiv \average{\calT \Phi_a(t,\bx)\Phi_b(t^{\prime},\bx^{\prime})}_{0},
 \qquad
 \Phi_{\eta}=\eta,\quad\Phi_{\varphi}=\varphi,
 \quad a,b\in\{\eta,\varphi\}.
 \label{app-eq:time-ordered-propagators}
\end{align}
Here, $\mathcal T$ denotes time ordering and $\average{\cdots}_{0}$ is the ground-state expectation value in the quadratic theory.
From the equations of motion, the propagators for each source index $b$ obey
\begin{align}
 -\partial_tG_{\varphi b}(t,\bx;t^{\prime},\bx^{\prime})-\calD^{n}_{x}G_{\eta b}(t,\bx;t^{\prime},\bx^{\prime})
 &=\delta_{\eta b}\,\delta(t-t')\delta^{(3)}(\bx-\bx'),
 \label{app-eq:density-Green-source}\\
\partial_tG_{\eta b}(t,\bx;t^{\prime},\bx^{\prime})-\calD^{\varphi}_{x}G_{\varphi b}(t,\bx;t^{\prime},\bx^{\prime})
 &=\delta_{\varphi b}\,\delta(t-t')\delta^{(3)}(\bx-\bx').
 \label{app-eq:phase-Green-source}
\end{align}
The subscript $x$ of the spatial operators represents the operators acting on the first argument $\bx$.
The propagators are obtained by solving these coupled equations with the boundary conditions from Eq.~\eqref{app-eq:boubdaey-cond} imposed on each argument.
The boundary conditions in the first argument are
\begin{align}
\lim_{z\to0^+}\frac{G_{\eta b}(t,\bx;t',\bx')}{2\sqrt{n_{\mathrm{GP}}(z)}}
=0,\qquad
\lim_{z\to0^+}\sqrt{n_{\mathrm{GP}}(z)}\,G_{\varphi b}(t,\bx;t',\bx')=0,
\label{app-eq:propagator-wall-first}
\end{align}
and the conditions in the second argument are
\begin{align}
\lim_{z'\to0^+}\frac{G_{a\eta}(t,\bx;t',\bx')}{2\sqrt{n_{\mathrm{GP}}(z')}}
=0,\qquad
\lim_{z'\to0^+}\sqrt{n_{\mathrm{GP}}(z')}\,G_{a\varphi}(t,\bx;t',\bx')=0.
\label{app-eq:propagator-wall-second}
\end{align}
Here $a,b\in\{\eta,\varphi\}$, and the other point is held fixed in the interior when either boundary limit is taken.

For reference, in an infinite homogeneous system with density $n_0$, the bulk propagator matrix in the field order $(\eta,\varphi)$ is given by
\begin{align}
\bm{G}^{\mathrm b}(\omega,\bk)=\frac{1}{\omega^2-E_k^2+\rmi0^+}
\begin{pmatrix}
n_0k^2/m & \rmi\omega\\
-\rmi\omega & g+k^2/(4mn_0)
\end{pmatrix},\qquad
E_k^2=c_s^2k^2+\frac{k^4}{4m^2},\quad c_s=\sqrt{\frac{gn_0}{m}}
\label{app-eq:homogeneous-propagators-reference}
\end{align}
with the Fourier convention
\begin{align}
\bm{G}^{\mathrm b}(\omega,\bk)=\int \rmd t\rmd^{3}\bx\, \rme^{+\rmi\omega(t-t')-\rmi\bk\cdot(\bx-\bx')}\bm{G}^{\mathrm b}(t,\bx;t^{\prime},\bx^{\prime}).
\end{align}
Here $k=|\bk|$, and $0^+$ represents an infinitesimal positive constant.
At zero frequency, the mixed components vanish and the diagonal components reduce to
\begin{align}
G^{\mathrm b}_{\eta\eta}(0,\bk)=-4m n_{0}\frac{1}{k^2+2\xi^{-2}},\qquad
G^{\mathrm b}_{\varphi\varphi}(0,\bk)=-\frac{m}{n_0}\frac{1}{k^2}.
\label{app-eq:homogeneous-static-reference}
\end{align}
The static density response is therefore short ranged, with exponential decay length $\xi/\sqrt{2}$, whereas the phase response has a long-range $1/k^2$ singularity.

\subsubsection{One-loop correction to the particle-number density}
The particle-number density is obtained from
\begin{align}
\average{\hat{n}(t,\bx)}
&=\frac{\int \calD\eta\calD\varphi\,n(t,\bx)\rme^{\rmi S_{\micro}}}
{\int \calD\eta\calD\varphi\,\rme^{\rmi S_{\micro}}}.
\end{align}
We evaluate this expectation value to one-loop order.
As we will see, the one-loop contribution provides a leading long-range correction at $z\gg \xi$.
The one-loop contribution from the interaction vertices comes from the cubic part $\calL_{\mathrm{int}}^{(3)}=-\eta(\nabla\varphi)^2/(2m)+\eta(\nabla [\eta/n_{\mathrm{GP}}(z)])^2/(8m)$, giving
\begin{align}
\average{\hat{n}(t,\bx)}
&=n_{\mathrm{GP}}(z)+\rmi\int_{z^{\prime}>0}\rmd t^{\prime}\rmd^{3}\bx^{\prime}\,
\average{\calT\eta(t,\bx)\calL_{\mathrm{int}}^{(3)}(t^{\prime},\bx^{\prime})}_{0,c}
+\text{(higher-loop terms)}.
\end{align}
The subscript $c$ represents contractions connected to the density insertion.
Here, we focus on the phase-vertex contribution from the first term of $\calL_{\mathrm{int}}$ in Eq.~\eqref{app-eq:inhomogeneous-interaction-L}:
\begin{align}
\delta n_{\mathrm{phase}}(z):=-\frac{\rmi}{2m}\int_{z^{\prime}>0}\rmd t^{\prime}\rmd^{3}\bx^{\prime}\,
\average{\calT\eta(t,\bx)\eta(t^{\prime},\bx^{\prime})
[\nabla_{\bx'}\varphi(t^{\prime},\bx^{\prime})]^2}_{0,c}.\label{app-eq:delN-phase}
\end{align}
The wall boundary conditions and the profile $n_{\mathrm{GP}}(z)$ break translation invariance in the $z$ direction, so that the contractions depend separately on $z$ and $z'$.
Integration over the vertex position $z'>0$ therefore leaves a dependence on the observation point's distance $z$ from the wall.
For the leading tail at $z\gg\xi$, the density-only cubic contribution is suppressed by $O((k\xi)^2)$ relative to the phase contribution.

We now evaluate the Wick contractions in Eq.~\eqref{app-eq:delN-phase}.
Time-translation invariance makes every contraction within the same vertex independent of the vertex time $t'$, as illustrated by
\begin{align}
G_{\varphi\varphi}(t',\bx';t'+\tau,\bx'')
&=G_{\varphi\varphi}(0,\bx';\tau,\bx''), \qquad \for\quad \tau>0.
\end{align}
The same relation holds for all propagator components.
Accordingly, the time integral acts only on the propagator connecting the density insertion to the vertex, selecting its zero-frequency component.
At zero frequency, the quadratic density and phase equations decouple because their coupling contains a time derivative, giving
\begin{align}
\int \rmd t'\,G_{\eta\varphi}(t,\bx;t',\bx')=0.
\end{align}
The two contractions connecting the density insertion to a phase field therefore vanish, leaving
\begin{align}
\delta n_{\mathrm{phase}}(z)
=\frac{\rmi}{2m}\int_{z^{\prime}>0}\rmd t^{\prime}\rmd^{3}\bx^{\prime}
G_{\eta\eta}(t,\bx;t^{\prime},\bx^{\prime})
\lim_{\tau\to 0^{+}}\lim_{\bx^{\prime\prime}\to \bx^{\prime}}
\left[\nabla_{\bx'}\!\cdot\!\nabla_{\bx''}
G_{\varphi\varphi}(0,\bx^{\prime};\tau,\bx^{\prime\prime})\right]
=
\scalebox{0.8}{
\begin{tikzpicture}[baseline=(b.base)]
\begin{feynhand}
	\vertex (b) at (0,-0.1) {};
	\vertex (v0) at (-2.4,0) {};
	\vertex at (-2.3,0.32) {\Large $z$};
	\vertex (v) at (-0.56,0) {};
	\vertex [dot] at (-0.56,0) {}; 
	\vertex at (-0.56,0.4) {\Large $z^{\prime}$}; 
	\propag [plain] (v0) to (v);
	\propag [sca] (v) to [out=45,in=-45,min distance=16mm] (v);
\end{feynhand}
\end{tikzpicture}}
.
\label{app-eq:phase-density-tadpole}
\end{align}
The diagram on the right represents this tadpole contribution, with the solid line denoting $G_{\eta\eta}$ and the dashed loop denoting $G_{\varphi\varphi}$.
The ultraviolet divergence of the phase-vertex contribution cancels against the density-only cubic contribution when the two vertices are combined.

\subsubsection{Effective boundary conditions at long distances}
We focus on the long-distance region $z\gg\xi$.
In this region, we approximate $n_{\mathrm{GP}}(z)$ by $n_0$ and call the resulting slowly varying phase solution $\varphi^{\mathrm{out}}$ the outer solution.
As the homogeneous propagators~\eqref{app-eq:homogeneous-static-reference} show, the density field has a short-range static response and can be eliminated in a long-wavelength expansion, leaving a description of the single phonon branch in terms of $\varphi$.
We determine the effective boundary condition for the continued outer field at $z=0$ by matching to the microscopic solution with the full profile $n_{\mathrm{GP}}(z)$.

To find the effective boundary condition, we consider modes proportional to $\rme^{-\rmi\omega t+\rmi\bk_\parallel\cdot\bx_\parallel}$, with amplitudes $\eta_{\omega,\bk_\parallel}(z)$ and $\varphi_{\omega,\bk_\parallel}(z)$, where $\bx_\parallel=(x,y)$ and $\bk_\parallel$ is the momentum parallel to the wall.
We choose $z_\ast$ in the overlap region where the GP profile is nearly uniform and the phonon field varies slowly:
\begin{align}
\xi\ll z_\ast\ll\min(k_\parallel^{-1},c_s/|\omega|).
\end{align}
The low-frequency, long-wavelength expansion is controlled by
\begin{align}
\epsilon\equiv\xi\sqrt{k_\parallel^2+\frac{\omega^2}{c_s^2}}\ll1.
\label{app-eq:boundary-expansion-parameter}
\end{align}

Our task is summarized as follows.
We first solve the microscopic coupled equations perturbatively in $\epsilon$ for $0<z\lesssim z_\ast$.
At $z=z_\ast$, we define the outer solution by matching its phase amplitude and derivative to the microscopic solution:
\begin{align}
\varphi^{\mathrm{out}}_{\omega,\bk_\parallel}(z_\ast)
=\varphi_{\omega,\bk_\parallel}(z_\ast),\qquad
\partial_z\varphi^{\mathrm{out}}_{\omega,\bk_\parallel}(z_\ast)
=\partial_z\varphi_{\omega,\bk_\parallel}(z_\ast).
\label{app-eq:phase-matching-conditions}
\end{align}
We then continue the matched outer solution to $z=0$ and express the matching conditions there as an effective boundary condition at $z=0$.

\paragraph{Inner solution.}
Since the propagators used in the one-loop diagram are defined by the quadratic action, it is sufficient to consider the linearized equations~\eqref{app-eq:density-linear-eom} and \eqref{app-eq:phase-linear-eom} for their mode functions.
For these Fourier modes, the two equations become
\begin{align}
\calD^n_{\bk_\parallel}\eta_{\omega,\bk_\parallel}(z)
\label{app-eq:boundary-density-response}
&=\rmi\omega\varphi_{\omega,\bk_\parallel}(z),\\
\partial_z\!\left[n_{\mathrm{GP}}(z)\partial_z\varphi_{\omega,\bk_\parallel}(z)\right]
&=k_\parallel^2n_{\mathrm{GP}}(z)\varphi_{\omega,\bk_\parallel}(z)+\rmi m\omega\eta_{\omega,\bk_\parallel}(z),
\label{app-eq:boundary-continuity}
\end{align}
with $\calD^n_{\bk_\parallel}\equiv\left.\calD^n\right|_{\nabla_\parallel\to\rmi\bk_\parallel}$.

At $\omega=k_\parallel=0$, Eq.~\eqref{app-eq:boundary-continuity} gives $n_{\mathrm{GP}}(z)\partial_z\varphi_{0,\bzero}(z)=C$ with a constant $C$.
Since $n_{\mathrm{GP}}(z)\sim z^2$, a nonzero $C$ would make the phase diverge as $z^{-1}$, contrary to Eq.~\eqref{app-eq:regular-wall-modes}.
The zero mode is therefore constant, $\varphi_{0,\bzero}(z)=\varphi_0\neq0$, with $C=0$.
We then solve the coupled equations perturbatively in $\omega\xi/c_s$ and $k_\parallel\xi$ about this zero mode $\varphi_{0}$.
We require the density response $\eta_{\omega,\bk_\parallel}(z)$ to satisfy the boundary condition~\eqref{app-eq:regular-wall-modes} and to match the phonon solution for the uniform background $n_{\mathrm{GP}}=n_0$ at $z\gg\xi$.
The overall amplitude is arbitrary, and for convenience we choose $\lim_{z\to0^+}\varphi_{\omega,\bk_\parallel}(z)=\varphi_0$ for all $\omega,\bk_\parallel$.

Differentiating the background equation~\eqref{app-eq:tree-density-equation} with respect to $\mu$ at fixed $g$ gives $\calD^n_{\bzero}\partial_\mu n_{\mathrm{GP}}=1$, so Eq.~\eqref{app-eq:boundary-density-response} gives $\eta_{\omega,\bk_\parallel}=\rmi\omega\varphi_0\partial_\mu n_{\mathrm{GP}}$ at leading order in $\epsilon$.
Eliminating $\eta$ from Eqs.~\eqref{app-eq:boundary-density-response} and \eqref{app-eq:boundary-continuity} leaves a phase equation that depends on $\omega$ and $\bk_\parallel$ only through $\omega^2$ and $k_\parallel^2$.
The first correction to the phase field is therefore second order, while the factor $\rmi\omega$ in the density response makes its next correction third order:
\begin{align}
\varphi_{\omega,\bk_\parallel}(z)
&=\varphi_0+O(\epsilon^2),
\label{app-eq:boundary-leading-phase}\\
\eta_{\omega,\bk_\parallel}(z)
&=\rmi\omega\varphi_0\,\partial_\mu n_{\mathrm{GP}}(z)
+O(\epsilon^3).
\label{app-eq:boundary-leading-density}
\end{align}
Since $\varphi_0$ is constant in $z$, the phase gradient starts at second order in $\epsilon$, i.e., $\partial_{z}\varphi_{\omega,\bk_\parallel}(z)=O(\epsilon^{2})$.

To determine the second-order behavior of this gradient, we substitute Eqs.~\eqref{app-eq:boundary-leading-phase} and \eqref{app-eq:boundary-leading-density} into the right-hand side of Eq.~\eqref{app-eq:boundary-continuity}:
\begin{align}
k_\parallel^2n_{\mathrm{GP}}(z)\varphi_{\omega,\bk_\parallel}(z)
+\rmi m\omega\eta_{\omega,\bk_\parallel}(z)
=\varphi_0\left[
k_\parallel^2n_{\mathrm{GP}}(z)
-m\omega^2\partial_\mu n_{\mathrm{GP}}(z)
\right]
+O(\epsilon^4).
\end{align}
Integrating Eq.~\eqref{app-eq:boundary-continuity} from $0$ to $z_\ast$, with $\lim_{z\to0^+}n_{\mathrm{GP}}(z)\partial_z\varphi_{\omega,\bk_\parallel}(z)=0$ and $n_{\mathrm{GP}}(z_\ast)\simeq n_0$, then gives
\begin{align}
\partial_z\varphi_{\omega,\bk_\parallel}(z_\ast)
\simeq\left[z_\ast\left(k_\parallel^2-\frac{\omega^2}{c_s^2}\right)
+\sqrt2\xi\left(\frac{\omega^2}{2c_s^2}-k_\parallel^2\right)\right]\varphi_0
+O(\epsilon^4).
\label{app-eq:boundary-matching-slope}
\end{align}
Hereafter, $\simeq$ denotes the omission of background corrections exponentially small in $z_\ast/\xi$.
We used the integrals
\begin{align}
\int_0^{z_\ast}\rmd z\,n_{\mathrm{GP}}(z)
\simeq n_0(z_\ast-\sqrt2\xi),\qquad
\int_0^{z_\ast}\rmd z\,\partial_\mu n_{\mathrm{GP}}(z)
\simeq\frac{1}{g}\left(z_\ast-\frac{\xi}{\sqrt2}\right).
\label{app-eq:boundary-profile-integrals}
\end{align}
The second integral follows by differentiating the first with respect to $\mu$ at fixed $g$ and $z_\ast$, with the use of $\partial_{\mu}n_{0}=1/g$ and $\partial_{\mu}\xi=-\xi/(2\mu)$.
The terms proportional to $z_\ast$ describe propagation in the uniform bulk, while those proportional to $\xi$ arise from the density variation near the wall.

\paragraph{Outer solution.}
We next consider the slowly varying outer solution $\varphi^{\mathrm{out}}$.
Setting $n_{\mathrm{GP}}=n_0$ in Eqs.~\eqref{app-eq:boundary-density-response} and \eqref{app-eq:boundary-continuity} gives the coupled equations for $\varphi^{\mathrm{out}}$ and the corresponding density amplitude $\eta^{\mathrm{out}}$:
\begin{align}
\left[g-\frac{\partial_z^2-k_\parallel^2}{4mn_0}\right]\eta^{\mathrm{out}}_{\omega,\bk_\parallel}(z)
&=\rmi\omega\varphi^{\mathrm{out}}_{\omega,\bk_\parallel}(z),\\
\frac{n_0}{m}\left(\partial_z^2-k_\parallel^2\right)\varphi^{\mathrm{out}}_{\omega,\bk_\parallel}(z)
&=\rmi\omega\eta^{\mathrm{out}}_{\omega,\bk_\parallel}(z).
\label{app-eq:outer-coupled-equations}
\end{align}
Combining these two equations, together with $c_s^2=gn_0/m$, yields
\begin{align}
\left[-\omega^2-c_s^2\left(\partial_z^2-k_\parallel^2\right)
+\frac{1}{4m^2}\left(\partial_z^2-k_\parallel^2\right)^2\right]
\varphi^{\mathrm{out}}_{\omega,\bk_\parallel}(z)=0.
\label{app-eq:outer-phase-equation}
\end{align}
For the slowly varying phonon solution in the homogeneous background, all spatial derivatives have the same order counting, $\partial_z,k_\parallel=O(\epsilon)$.
Then, the operator $(\partial_z^2-k_\parallel^2)^2$ is fourth order in $\epsilon$, so that the equation reduces to
\begin{align}
\partial_z^2\varphi^{\mathrm{out}}_{\omega,\bk_\parallel}(z)
=\left[k_\parallel^2-\frac{\omega^2}{c_s^2}+O(\epsilon^4)\right]\varphi^{\mathrm{out}}_{\omega,\bk_\parallel}(z).
\label{app-eq:outer-phase-solution}
\end{align}

\paragraph{Matching.}
We extrapolate the outer field $\varphi^{\mathrm{out}}_{\omega,\bk_\parallel}$, defined for $z\geq z_\ast$, to $0\leq z\leq z_\ast$ using Eq.~\eqref{app-eq:outer-phase-solution}.
Its value and derivative at $z_\ast$ are fixed by the matching conditions~\eqref{app-eq:phase-matching-conditions}, so that this continuation determines an effective boundary condition at $z=0$.
The inner solution gives $\varphi_{\omega,\bk_\parallel}(z)=\varphi_0+O(\epsilon^2)$ and $\partial_z\varphi_{\omega,\bk_\parallel}(z)=O(\epsilon^2)$.
Evaluating these results at $z=z_\ast$ and applying the matching conditions~\eqref{app-eq:phase-matching-conditions} gives
\begin{align}
\varphi^{\mathrm{out}}_{\omega,\bk_\parallel}(z_\ast)=\varphi_0+O(\epsilon^2),\qquad
\partial_z\varphi^{\mathrm{out}}_{\omega,\bk_\parallel}(z_\ast)=O(\epsilon^2).
\label{app-eq:matching-zast}
\end{align}

At zeroth order in $\epsilon$, Eq.~\eqref{app-eq:outer-phase-solution} gives $\partial_z^2\varphi^{\mathrm{out}}=0$, and Eq.~\eqref{app-eq:matching-zast} selects $\varphi^{\mathrm{out}}=\varphi_0$ throughout $0\leq z\leq z_\ast$.
At second order in $\epsilon$, the right-hand side of Eq.~\eqref{app-eq:outer-phase-solution} is $(k_\parallel^2-\omega^2/c_s^2)\varphi_0$, so that integrating Eq.~\eqref{app-eq:outer-phase-solution} from $z$ to $z_\ast$ yields
\begin{align}
\partial_z\varphi^{\mathrm{out}}_{\omega,\bk_\parallel}(z)
=\partial_z\varphi^{\mathrm{out}}_{\omega,\bk_\parallel}(z_\ast)
+(z-z_\ast)\left(k_\parallel^2-\frac{\omega^2}{c_s^2}\right)\varphi_0+O(\epsilon^4).
\label{app-eq:outer-phase-extrapolated}
\end{align}
Substituting the matching condition~\eqref{app-eq:phase-matching-conditions} and the derivative of the inner solution~\eqref{app-eq:boundary-matching-slope} into Eq.~\eqref{app-eq:outer-phase-extrapolated} at $z=0$ gives
\begin{align}
\partial_z\varphi^{\mathrm{out}}_{\omega,\bk_\parallel}(0)
&=\partial_z\varphi_{\omega,\bk_\parallel}(z_\ast)
-z_\ast\left(k_\parallel^2-\frac{\omega^2}{c_s^2}\right)\varphi_0+O(\epsilon^4) \nonumber\\
&\simeq\sqrt{2}\xi\left(\frac{\omega^2}{2c_s^2}-k_\parallel^2\right)\varphi_0+O(\epsilon^4).
\label{app-eq:outer-matching-slope}
\end{align}
The terms proportional to $z_\ast$ cancel, leaving the contribution from the surface layer.
Equation~\eqref{app-eq:outer-phase-extrapolated} shows that $\partial_z\varphi^{\mathrm{out}}_{\omega,\bk_\parallel}(z)=O(\epsilon^2)$ throughout $0\leq z\leq z_\ast$, so that integration from $0$ to $z_\ast$ yields
\begin{align}
\varphi^{\mathrm{out}}_{\omega,\bk_\parallel}(0)
=\varphi^{\mathrm{out}}_{\omega,\bk_\parallel}(z_\ast)
-\int_0^{z_\ast}\rmd z\,\partial_z\varphi^{\mathrm{out}}_{\omega,\bk_\parallel}(z)
=\varphi_0+O(\epsilon^2).
\label{app-eq:outer-phase-expanded}
\end{align}
This equation allows for replacing $\varphi_{0}$ on the right-hand side of Eq.~\eqref{app-eq:outer-matching-slope} with $\varphi^{\mathrm{out}}_{\omega,\bk_\parallel}(0)$.
As a result, with exponentially small background corrections omitted, we obtain
\begin{align}
\left.\left[\partial_z-\sqrt{2}\xi\left(\frac{\omega^2}{2c_s^2}-k_\parallel^2\right)\right]\varphi^{\mathrm{out}}_{\omega,\bk_\parallel}(z)\right|_{z=0}
=O\left(\epsilon^4\right).
\label{app-eq:matched-phase-boundary}
\end{align}
Neglecting the fourth-order remainder gives the effective boundary condition through second order in frequency and momentum.
The term proportional to $\xi$ arises from the density variation in the surface layer and contains one additional derivative relative to $\partial_z$.
At leading derivative order, this term is omitted, giving the Neumann condition for the continued outer field:
\begin{align}
\left.\partial_z\varphi^{\mathrm{out}}_{\omega,\bk_\parallel}(z)\right|_{z=0}=0.\label{app-eq:Neumann-BC}
\end{align}

\subsubsection{Phonon contribution to the long-distance density profile}
The effective Neumann condition~\eqref{app-eq:Neumann-BC} makes the propagators depend separately on $z$ and $z'$, even where $n_{\mathrm{GP}}(z)\simeq n_0$.
We evaluate the propagators of the continued bulk fields at leading order for $k\xi\ll1$ and $|\omega|\xi/c_s\ll1$, as appropriate for $z,z'\gg\xi$.
We use the partial Fourier transform defined by
\begin{align}
G_{ab}(t,\bx;t',\bx')
=\int\frac{\rmd\omega}{2\pi}\frac{\rmd^2\bk_\parallel}{(2\pi)^2}
\rme^{-\rmi\omega(t-t')+\rmi\bk_\parallel\cdot(\bx_\parallel-\bx'_\parallel)}
G_{ab}(\omega,\bk_\parallel;z,z').
\end{align}
At this order, the density operator~\eqref{app-eq:density-operator} reduces to $\calD^n=g$.
Then, Eq.~\eqref{app-eq:density-Green-source} and the corresponding source equation in the second argument give
\begin{align}
gG_{\eta b}(\omega,\bk_\parallel;z,z')
&=\rmi\omega G_{\varphi b}(\omega,\bk_\parallel;z,z')-\delta_{\eta b}\delta(z-z'),\\
gG_{a\eta}(\omega,\bk_\parallel;z,z')
&=-\rmi\omega G_{a\varphi}(\omega,\bk_\parallel;z,z')-\delta_{a\eta}\delta(z-z').
\end{align}
Thus, all components containing $\eta$ follow from the phase propagator:
\begin{align}
G_{\eta\varphi}
=\frac{\rmi\omega}{g}G_{\varphi\varphi},\qquad
G_{\varphi\eta}=-\frac{\rmi\omega}{g}G_{\varphi\varphi},
\qquad G_{\eta\eta}
=\frac{\omega^2}{g^2}G_{\varphi\varphi}-\frac{1}{g}\delta(z-z').
\label{app-eq:outer-density-reconstruction}
\end{align}
All propagators in Eq.~\eqref{app-eq:outer-density-reconstruction} have the same arguments $(\omega,\bk_\parallel;z,z')$.
No independent boundary condition for $\eta$ is needed at this order, since its propagators are fixed by $G_{\varphi\varphi}$.

At zero frequency, Eq.~\eqref{app-eq:outer-density-reconstruction} gives $G_{\eta\eta}(0,\bk_\parallel;z,z')=-\delta(z-z')/g$, entirely from the contact term.
Fourier transforming in $\bk_\parallel$ therefore yields the static density response in the half-space:
\begin{align}
\int\rmd t'\,G_{\eta\eta}(t,\bx;t',\bx')=-\frac{1}{g}\delta^{(3)}(\bx-\bx').
\label{app-eq:outer-static-density-response}
\end{align}
At this order, the static density propagator is local, as in the homogeneous system, and does not itself generate a power-law long-distance tail.

Substituting $G_{\eta\varphi}=(\rmi\omega/g)G_{\varphi\varphi}$ into Eq.~\eqref{app-eq:phase-Green-source} gives
\begin{align}
\left[\omega^2+c_s^2(\partial_z^2-k_\parallel^2)\right]
G_{\varphi\varphi}(\omega,\bk_\parallel;z,z')=g\delta(z-z').
\end{align}
The effective Neumann condition~\eqref{app-eq:Neumann-BC} imposes the following conditions on $G_{\varphi\varphi}$ in its two arguments:
\begin{align}
\left.\partial_zG_{\varphi\varphi}(\omega,\bk_\parallel;z,z')\right|_{z=0}=0,\qquad
\left.\partial_{z'}G_{\varphi\varphi}(\omega,\bk_\parallel;z,z')\right|_{z'=0}=0.
\label{app-eq:outer-phase-propagator-BC}
\end{align}
In each condition, the other endpoint is held fixed in $z>0$.
At mean-field level for a weakly interacting Bose gas, $\chi=1/g$, so that the canonically normalized field is $\pi=\varphi/\sqrt{g}$.
Its propagator is
\begin{align}
\rmi G(t,\bx;t',\bx')\equiv\average{\calT\pi(t,\bx)\pi(t',\bx')}_{0}=\frac{\rmi}{g}G_{\varphi\varphi}(t,\bx;t',\bx').
\end{align}
This is the propagator defined in the main text for $d=3$.

We now evaluate the density correction $\delta n_{\mathrm{phase}}(z)$ in Eq.~\eqref{app-eq:phase-density-tadpole}.
Using Eq.~\eqref{app-eq:outer-static-density-response} and performing the spatial integral then gives
\begin{align}
\delta n_{\mathrm{phase}}(z)=-\frac{1}{2m}
\lim_{\tau\to0^+}\lim_{\bx'\to\bx}
\nabla_{\bx}\!\cdot\!\nabla_{\bx'}
\left[\rmi G(0,\bx;\tau,\bx')\right].
\label{app-eq:local-phase-tadpole}
\end{align}
For the mean-field equation of state $p(\mu)=\mu^2/(2g)$, $\chi'=0$, so that Eq.~\eqref{app-eq:local-phase-tadpole} is Eq.~\eqref{eq:limit-dn} specialized to the three-dimensional weakly interacting Bose gas.

\clearpage
\section{\texorpdfstring{Detailed calculation of $\delta n_{\tail}(z)$}{Detailed calculation of the density tail}}
We now evaluate the zero-temperature tail for a general superfluid in $d$ spatial dimensions.
The image term in Eq.~\eqref{eq:image-decomposition} gives
\begin{align}
\delta n_{\tail}(z)
=\lim_{t'\to t^+}\lim_{\bx'\to\bx}
\biggl[
-\frac{\nabla_{\bx}\!\cdot\!\nabla_{\bx'}}{2m}
+\frac{\chi'}{2\chi}\partial_t\partial_{t'}
\biggr]\rmi G_{\mathrm{im}}(t,\bx;t',\bx').
\label{app-eq:density-tail-from-phase}
\end{align}
For the canonically normalized field, the image propagator is
\begin{align}
G_{\mathrm{im}}(t,\bx;t',\bx')
&=\int\frac{\rmd\omega}{2\pi}\frac{\rmd^d\bk}{(2\pi)^d}
\frac{\rme^{-\rmi\omega(t-t')+\rmi\bk_\parallel\cdot(\bx_\parallel-\bx'_\parallel)+\rmi k_z(z+z')}}
{\omega^2-c_s^2k^2+\rmi0^+}.
\label{app-eq:real-time-phase-image}
\end{align}
The time derivatives are taken before setting $t=0$ and $t'=\tau>0$ and sending $\tau\to0^+$.

We first consider the spatial derivative term in Eq.~\eqref{app-eq:density-tail-from-phase}.
The Feynman prescription gives
\begin{align}
\rmi\int\frac{\rmd\omega}{2\pi}
\frac{\rme^{\rmi\omega\tau}}{\omega^2-c_s^2k^2+\rmi0^+}
=\frac{\rme^{-\rmi c_s k\tau}}{2c_s k},\qquad \tau>0.
\end{align}
Let $\bm r=(\bx_\parallel-\bx'_\parallel,z+z')$ and $r=|\bm r|$.
For $d>1$, while the momentum integral of this result has a UV divergence, one can correctly read off its finite long-distance behavior by introducing an appropriate convergence factor as~\cite{Berestetskii:1982qed}
\begin{align}
\lim_{\tau\to0^+}\rmi G_{\mathrm{im}}(0,\bx;\tau,\bx')
=\frac{1}{2c_s}\lim_{\alpha\to0^+}
\int\frac{\rmd^d\bk}{(2\pi)^d}\frac{\rme^{\rmi\bk\cdot\bm r-\alpha k}}{k}.
\label{app-eq:equal-time-phase-image}
\end{align}
The Fourier-transform identity takes the following form in $d$ dimensions:
\begin{align}
\lim_{\alpha\to0^+}\int\frac{\rmd^d\bk}{(2\pi)^d}
k^\nu\rme^{\rmi\bk\cdot\bm r-\alpha k}
=\frac{2^\nu\Gamma((d+\nu)/2)}
{\pi^{d/2}\Gamma(-\nu/2)\,r^{d+\nu}},\qquad
r>0,\quad d+\nu>0.
\end{align}
This gives, for $\nu=-1$ and $d>1$,
\begin{align}
\lim_{\tau\to0^+}\rmi G_{\mathrm{im}}(0,\bx;\tau,\bx')
=\frac{\Gamma((d-1)/2)}{4\pi^{(d+1)/2}c_s\,r^{d-1}}.
\end{align}

For $d>1$, taking one spatial derivative at each endpoint before setting $\bx'=\bx$ gives
\begin{align}
\lim_{\tau\to0^+}\lim_{\bx'\to\bx}
\nabla_{\bx}\!\cdot\!\nabla_{\bx'}
\left[\rmi G_{\mathrm{im}}(0,\bx;\tau,\bx')\right]
=\frac{(2d-1)\Gamma((d+1)/2)}
{2^{d+2}\pi^{(d+1)/2}c_s z^{d+1}}.
\label{app-eq:phase-gradient-image}
\end{align}
The parallel derivatives act on coordinate differences and contribute $(d-1)^2$, while the normal derivatives act on $z+z'$ and contribute $d(d-1)$ before simplifying the Gamma functions.

We next consider the temporal derivative term in Eq.~\eqref{app-eq:density-tail-from-phase}.
Taking the time derivatives contributes a factor $(c_s k)^2$ to the spectral kernel and gives
\begin{align}
\lim_{\tau\to0^+}\lim_{\bx'\to\bx}
\left.\partial_t\partial_{t'}
\left[\rmi G_{\mathrm{im}}(t,\bx;t',\bx')\right]
\right|_{t=0,\,t'=\tau}=\frac{c_s}{2}\lim_{\alpha\to0^+}
\int\frac{\rmd^d\bk}{(2\pi)^d}
k\,\rme^{2\rmi k_z z-\alpha k}
=-\frac{c_s\Gamma((d+1)/2)}
{2^{d+2}\pi^{(d+1)/2}z^{d+1}}.
\label{app-eq:phase-time-image}
\end{align}
The derivatives are evaluated at $\tau>0$, so that no time-ordering contact term enters.

For $d=1$, we avoid the infrared divergence of the undifferentiated phase propagator by differentiating its spectral kernel before integrating over momentum.
The resulting contractions are again given by Eqs.~\eqref{app-eq:phase-gradient-image} and \eqref{app-eq:phase-time-image}, evaluated at $d=1$.

Substituting the spatial and temporal contractions into Eq.~\eqref{app-eq:density-tail-from-phase} yields
\begin{align}
\delta n_{\tail}(z)
=-\frac{\Gamma((d+1)/2)}
{2^{d+3}\pi^{(d+1)/2}z^{d+1}}
\left(\frac{2d-1}{m c_s}+c_s\frac{\chi'}{\chi}\right).
\label{app-eq:general-zeroT-tail}
\end{align}
This reproduces Eq.~\eqref{eq:general_zeroT} in the main text.

%%%%%%%%%%%%%%%%%%%%%%%%%%%%%%%%%%%%%%%%%%%%%%%%%%%%%%%%%%%%%%%%%%%%%%%%%%%%%%%%%%%%%%%
%%%%%%%%%%%%%%%%%%%%%%%%%%%%%%%%%%%%%%%%%%%%%%%%%%%%%%%%%%%%%%%%%%%%%%%%%%%%%%%%%%%%%%%
%%%%%%%%%%%%%%%%%%%%%%%%%%%%%%%%%%%%%%%%%%%%%%%%%%%%%%%%%%%%%%%%%%%%%%%%%%%%%%%%%%%%%%%
%%%%%%%%%%%%%%%%%%%%%%%%%%%%%%%%%%%%%%%%%%%%%%%%%%%%%%%%%%%%%%%%%%%%%%%%%%%%%%%%%%%%%%%
%%%%%%%%%%%%%%%%%%%%%%%%%%%%%%%%%%%%%%%%%%%%%%%%%%%%%%%%%%%%%%%%%%%%%%%%%%%%%%%%%%%%%%%
%%%%%%%%%%%%%%%%%%%%%%%%%%%%%%%%%%%%%%%%%%%%%%%%%%%%%%%%%%%%%%%%%%%%%%%%%%%%%%%%%%%%%%%
\end{document}